\documentclass[conference]{IEEEtran}
\usepackage{subfigure}
\usepackage{color}
\usepackage{cite}
\usepackage{amsmath}
\usepackage{cite}
\usepackage{graphicx}
\usepackage{epstopdf}
\usepackage{amsfonts,amsmath,amssymb}
\usepackage{graphicx}
\usepackage{url}
\usepackage{bm}
\usepackage{bbm}
\usepackage{subfigure}
\usepackage{stfloats}
\usepackage{color}

\usepackage{cite}

\DeclareMathOperator*{\argmax}{argmax}
\DeclareMathOperator*{\argmin}{argmin}

\newcommand{\Hnull}{\mathcal{H}_0}
\newcommand{\Halt}{\mathcal{H}_1}

\newcommand{\Honull}{\mathcal{{D}}_0}
\newcommand{\Hoalt}{\mathcal{{D}}_1}

\usepackage{cite,graphicx,amsmath,amssymb,cite,algorithm}

\begin{document}
\newtheorem{theorem}{\textbf{Theorem}}
\newtheorem{proposition}{\textbf{Proposition}}
\newtheorem{corollary}{\textbf{Corollary}}
\newtheorem{remark}{Remark}

\title{Covert Communication in Wireless Relay Networks}
\author{
\IEEEauthorblockN{Jinsong~Hu$^{\ast}$, Shihao~Yan$^{\dag}$, Xiangyun~Zhou$^{\dag}$, Feng~Shu$^{\ast}$, and Jiangzhou Wang$^{\ddagger}$}
\IEEEauthorblockA{$^{\ast}$School of Electronic and Optical Engineering, Nanjing University of Science and Technology, Nanjing, Jiangsu, China}
\IEEEauthorblockA{$^{\dag}$Research School of Engineering, The Australian National University, Canberra, ACT, Australia}
\IEEEauthorblockA{$^{\ddagger}$School of Engineering and Digital Arts, University of Kent, Canterbury, Kent, U.K.}
\IEEEauthorblockA{Emails:\{jinsong\_hu, shufeng\}@njust.edu.cn,~\{shihao.yan, xiangyun.zhou\}@anu.edu.au,~j.z.wang@kent.ac.uk}
}

\maketitle
\begin{abstract}
Covert communication aims to shield the very existence of wireless transmissions in order to guarantee a strong security in wireless networks. In this work, for the first time we examine the possibility and achievable performance of covert communication in one-way relay networks. Specifically, the relay opportunistically transmits its own information to the destination covertly on top of forwarding the source's message, while the source tries to detect this covert transmission to discover the illegitimate usage of the recourse (e.g., power, spectrum) allocated only for the purpose of forwarding source's information. The necessary condition that the relay can transmit covertly without being detected is identified and the source's detection limit is derived in terms of the false alarm and miss detection rates. Our analysis indicates that boosting the forwarding ability of the relay (e.g., increasing its maximum transmit power) also increases its capacity to perform the covert communication in terms of achieving a higher effective covert rate subject to some specific requirement on the source's detection performance.
\end{abstract}

\section{Introduction}

Security and privacy are critical in existing and future wireless networks since a large amount of confidential information (e.g., location, credit card information, physiological information for e-health) is transferred over the open wireless medium \cite{bloch2011physical}. Against this background, conventional cryptography and information-theoretic secrecy technologies have been developed to offer progressively higher levels of security by protecting the content of the message against eavesdropping \cite{yan2016artificial,hu2016robust,hu2017artificial}. However, these technologies cannot mitigate the threat to a user's security and privacy from discovering the presence of the user or communication. Meanwhile, this strong security (i.e., hiding the wireless transmission) is desired in many application scenarios of wireless communications, such as covert military operations and avoiding to be tracked in vehicular ad hoc networks. As such, the hiding of communication termed covert communication or low probability of detection communication, which aims to shield the very existence of wireless transmissions against a warden to achieve security, has recently drawn significant research interests and is emerging as a cutting-edge technique in the context of wireless communication security\cite{bash2015hiding,bloch2016covert}.

The fundamental limit of covert communication has been studied under various channel conditions, such as additive white Gaussian noise (AWGN) channel \cite{bash2013limits}, binary symmetric channel \cite{pak2013reliable}, and discrete memoryless channel \cite{wang2016fundamental}. It is proved that $\mathcal{O}(\sqrt{n})$ bits of information can be transmitted to a legitimate receiver reliably and covertly in $n$ channel uses as $n \rightarrow \infty$. This means that the associated covert rate is zero due to $\lim_{n\rightarrow\infty}\mathcal{O}(\sqrt{n})/n\rightarrow0$. Following these pioneering works on covert communication, a positive rate has been proved to be achievable when the warden has uncertainty on his receiver noise power \cite{lee2015achieving,goeckel2016covert}, the warden does not know when the covert communication happens \cite{bash2014LPD}, or an uniformed jammer comes in to help \cite{Sobers2015Covert}. Most recently, \cite{BiaoHe2017on} has examined the impact of noise uncertainty on covert communication  by considering two practical uncertainty models in order to debunk the myth of this impact. In addition, the effect of the imperfect channel state information and finite blocklength (i.e., finite $n$) on covert communication has been investigated in \cite{Shahzad2017Covert} and \cite{ShihaoYan2017Covert}, respectively.

In this work, for the first time we consider covert communication in the context of one-way relay networks. This is motivated by the scenario where the relay (R) tries to transmit its own information to the destination (D) on top of forwarding the information from the source (S) to D, while S forbids R's transmission of its own message since the resource (e.g., power, spectrum) allocated to R by S is dedicated to be solely used on aiding the transmission from S to D. As such, R's transmission of its own message to D should be kept covert from S, where S acts as the warden trying to detect this covert communication. We first identify the necessary condition that the covert transmission from R to D can possibly occur without being detected by S and then derive the detection limit at S in terms of the false alarm and miss detection rates under this condition. In addition, we analyze the achievable effective covert rate subject to a requirement on the detection performance at S. Our examination demonstrates a tradeoff between R's ability to aid the transmission from S to D and R's capability to conduct the covert communication.

\section{System Model}

\subsection{Considered Scenario and Adopted Assumptions}

\begin{figure}
  \centering
  \includegraphics[scale=0.64]{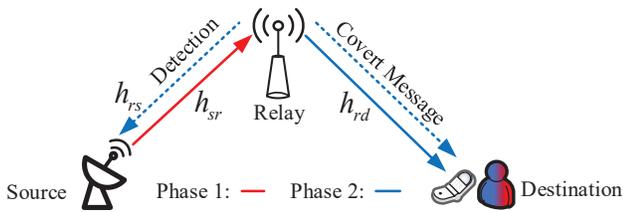}\\
  \caption{Covert communication in one-way relay networks.}\label{fig1}
\end{figure}

As shown in Fig.~\ref{fig1}, in this work we consider a one-way relay network, in which S transmits information to D with the aid of R, since a direct link from S to D is not available. As mentioned in the introduction, S allocates some resource to R in order to seek its help to relay the message to D. However, in some scenarios R may intend to use this resource to transmit its own message to D as well, which is forbidden by S and thus should be kept covert from S. As such, in the considered system model S is also the warden to detect whether R transmits its own information to D when it is aiding the transmission from S to D.

We assume the wireless channels within our system model are subject to independent quasi-static Rayleigh fading with equal block length and the channel coefficients are independent and identically distributed (i.i.d.) circularly symmetric complex Gaussian random variables with zero-mean and unit-variance. We also assume that each node is equipped with a single antenna. The channel from S to R is denoted by $h_{sr}$ and the channel from R to D is denoted by $h_{rd}$. We assume R knows both $h_{sr}$ and $h_{rd}$ perfectly, while S only knows $h_{sr}$ and D only knows $h_{rd}$. Considering channel reciprocity, we assume the channel from R to S (denoted by $h_{rs}$) is the same as $h_{sr}$ and thus perfectly known by S. We further assume that R operates in the half-duplex mode and thus the transmission from S to D occurs in two phases: phase 1 (S transmits to R) and phase 2 (R transmits to D).

\subsection{Transmission from Source to Relay (Phase 1)}

In phase 1, the received signal at R is given by
\begin{align}
\mathbf{y}_r[i]=\sqrt{P_s}h_{sr}\mathbf{x}_b[i]+\mathbf{n}_r[i],
\end{align}
where $P_s$ is the transmit power of source, $\mathbf{x}_b$ is the transmitted signal by S satisfying $\mathbb{E}[\mathbf{x}_b[i]\mathbf{x}^{\dag}_b[i]]=1$, $i = 1, 2, \dots, n$ is the index of each channel use ($n$ is the total number of channel uses in each phase), and $\mathbf{n}_r[i]$ is the AWGN at relay with $\sigma^2_r$ as its variance, i.e., $\mathbf{n}_r[i] \thicksim\mathcal{CN}(0,\sigma^2_r)$.

In this work, we consider that R operates in the amplify-and-forward mode and thus R will forward a linearly amplified version of the received signal to D in phase 2. As such, the forwarded (transmitted) signal by R is given by
\begin{align}\label{x_r}
\mathbf{x}_r[i]=G\mathbf{y}_r[i]=G(\sqrt{P_s}h_{sr}\mathbf{x}_b[i]+\mathbf{n}_r[i]),
\end{align}
where the received signal is scaled by a scalar $G$. In order to guarantee the power constraint at R, the value of G must be chosen such that $\mathbb{E}[\mathbf{x}_r[i]\mathbf{x}^{\dag}_r[i]]=1$, which leads to $G=1/\sqrt{P_s|h_{sr}|^2+\sigma^2_r}$.

In this work, we consider a fixed-rate transmission from S to D, in which this rate is denoted by $R_{sd}$. We also consider a maximum power constraint at R, i.e., $P_r \leq P_r^{\mathrm{max}}$. As such, although R knows both $h_{sr}$ and $h_{rd}$ perfectly, transmission outage from S to D still incurs when $C_{sd}^{\ast}<R_{sd}$, where $C_{sd}^{\ast}$ is the channel capacity from S to D for $P_r = P_r^{\mathrm{max}}$. Then, the transmission outage probability is given by $\delta = \mathcal{P}[C_{sd}^{*}<R_{sd}]$. In practice, the pair of $R_{sd}$ and $\delta$ determines the specific aid (i.e., the value of $P_r^{\mathrm{max}}$) required by S from R, which relates to the amount of resource allocated to R by S as a payback. In this work, we assume both $R_{sd}$ and $\delta$ are predetermined, which leads to a predetermined $P_r^{\mathrm{max}}$.

\section{Transmission Strategies at Relay}

In this section, we detail the transmission strategies of R when it does and does not transmit its own information to D. We also determine the condition that R can transmit its own message to D without being detected by S with probability one, in which the probability to guarantee this condition is also derived.

\subsection{Relay's Transmission without Covert Message}

In the case when relay does not transmit its own message (i.e., covert message) to Bob, it only transmit $\mathbf{x}_r$ to D. Accordingly, the received signal at D is given by
\begin{align}
\mathbf{y}_d[i]&=\sqrt{P_r^0}h_{rd}\mathbf{x}_r[i]+\mathbf{n}_d[i]  \\
&=\sqrt{P_r^0}Gh_{rd}\sqrt{P_s}h_{sr}\mathbf{x}_b[i]+\sqrt{P_r^0}Gh_{rd}\mathbf{n}_r[i]+\mathbf{n}_d[i], \notag
\end{align}
where $P_r^0$ is the transmit power of $\mathbf{x}_r$ at R in this case and $\mathbf{n}_d[i]$ is the AWGN at D with $\sigma^2_d$ as the variance, i.e., $\mathbf{n}_d[i]\thicksim\mathcal{CN}(0,\sigma^2_d)$. Accordingly, the signal-to-noise ratio (SNR) at the destination for $\mathbf{x}_b$ is given by
\begin{align} \label{gamma_d1}
\gamma_d&=\frac{P_s|h_{sr}|^2P_r^0|h_{rd}|^2 G^2}{P_r^0|h_{rd}|^2G^2\sigma^2_r+\sigma^2_d} =\frac{\gamma_{1}\gamma_{2}}{\gamma_{1}+\gamma_{2}+1},
\end{align}
where $\gamma_{1}\triangleq(P_s|h_{sr}|^2)/\sigma^2_r$ and $\gamma_{2}\triangleq(P_r^0|h_{rd}|^2)/\sigma^2_d$.

For the fixed-rate transmission, R does not have to adopt the maximum transmit power for each channel realization in order to guarantee a specific transmission outage probability. When the transmission outage occurs (i.e., $C_{sd}^{*}<R_{sd}$ occurs), R will not transmit (i.e., $P_r^0 = 0$). When $C_{sd}^{*} \geq R_{sd}$, R only has to ensure $C_{sd} = R_{sd}$, where $C_{sd} = {1}/{2}\log_2(1+\gamma_d)$. Then, following \eqref{gamma_d1} the transmit power of R when $C_{sd}^{*} \geq R_{sd}$ is given by $P_r^{0}={\mu\sigma^2_d}/{|h_{rd}|^2}$, where
\begin{align}\label{mu}
\mu\triangleq\frac{(P_s|h_{sr}|^2+\sigma^2_r)(2^{2R_{sd}}-1)}{\left[P_s|h_{sr}|^2-\sigma^2_r(2^{2R_{sd}}-1)\right]}.
\end{align}
Noting $\gamma_d < \gamma_1$, we have $1/2\log_2(1+\gamma_1)>R_{sd}$ when $C_{sd} = R_{sd}$. As such, $\mu$ given in \eqref{mu} is nonnegative.
Following \eqref{gamma_d1}, we note that $C_{sd}^{*} \geq R_{sd}$ requires $|h_{rd}|^2 \geq {\mu\sigma^2_d}/{P_r^{\mathrm{max}}}$. As such, the transmit power of R without covert message is given by
\begin{align}\label{PR0}
P_r^{0}=\left\{
  \begin{array}{ll}
    \frac{\mu\sigma^2_d}{|h_{rd}|^2}, & |h_{rd}|^2 \geq \frac{\mu\sigma^2_d}{P_r^{\mathrm{max}}}, \\
    0,  &|h_{rd}|^2 <\frac{\mu\sigma^2_d}{P_r^{\mathrm{max}}}.
  \end{array}
\right.
\end{align}

\subsection{Relay's Transmission with Covert Message}

In the case when R transmits the covert message to D on top of forwarding $\mathbf{x}_b$, the received signal at D is given by
\begin{align}\label{YD2}
\mathbf{y}_d[i]&=\sqrt{P_r^1}Gh_{rd}\sqrt{P_s}h_{sr}\mathbf{x}_b[i]+\sqrt{P_{\Delta}}h_{rd}\mathbf{x}_c[i] \notag \\
&~~~+\sqrt{P_r^1}Gh_{rd}\mathbf{n}_r[i]+\mathbf{n}_d[i].
\end{align}
where $P_r^1$ is the transmit power of R to forward $\mathbf{x}_b$ under this case and $P_{\Delta}$ is the power of R used to transmit the covert message $\mathbf{x}_c$ satisfying $\mathbb{E}[\mathbf{x}_c[i]\mathbf{x}^{\dag}_c[i]]=1$. In this work, we assume that $P_{\Delta}$ is fixed for all channel realizations.
In general, the transmit power of a covert message is significantly less than that of the forwarded message, i.e., $P_{\Delta} << P_r^1$. As such, here we assume D always first decodes $\mathbf{x}_b$ with $\mathbf{x}_c$ as interference. Following \eqref{YD2}, the signal-to-interference-plus-noise ratio (SINR) for $\mathbf{x}_b$ is given by
\begin{align}\label{gamma_d2}
\gamma_d&=\frac{P_s|h_{sr}|^2P_r^1|h_{rd}|^2 G^2}{P_r^1|h_{rd}|^2G^2\sigma^2_r+P_{\Delta}|h_{rd}|^2+\sigma^2_d} \notag \\
&=\frac{\gamma_{1}\gamma_{3}}{\gamma_{3}+(\gamma_{1}+1)\left(\gamma_{3}P_{\Delta}/P_r^1+1\right)},
\end{align}
where $\gamma_{3}\triangleq(P_r^1|h_{rd}|^2)/\sigma^2_d$. Then, when $C_{sd} = R_{sd}$ we have
\begin{align}\label{Pr1_special}
P_r^{1}=\mu P_{\Delta}+\frac{\mu\sigma^2_d}{|h_{rd}|^2},
\end{align}
which requires $C_{sd}^{*} \geq R_{sd}$ that leads to $ |h_{rd}|^2 \geq {\mu\sigma^2_d}/{[P_r^{\mathrm{max}}-(\mu+1)P_{\Delta}]}$. Considering the maximum power constraint at R (i.e., $P_r^1 + P_{\Delta} \leq P_r^{\mathrm{max}}$ under this case), R has to give up the transmission of the covert message (i.e., $P_{\Delta} = 0$) when $P_r^1>P_r^{\mathrm{max}}-P_{\Delta}$ and sets $P_r^1$ the same as $P_r^0$ given in  \eqref{PR0}. This is due to the fact that S knows $h_{rs}$ and it can detect with probability one when the total transmit power of R is greater than $P_r^{\mathrm{max}}$. Then, the transmit power of $\mathbf{x}_r$ under this case is given by
\begin{align}\label{PR1}
&P_r^{1}=\\
&\left\{
  \begin{array}{ll}
    \mu P_{\Delta}+\frac{\mu\sigma^2_d}{|h_{rd}|^2}, & |h_{rd}|^2 \geq \frac{\mu\sigma^2_d}{P_r^{\mathrm{max}}-(\mu+1)P_{\Delta}}, \\
    \frac{\mu\sigma^2_d}{|h_{rd}|^2}, & \frac{\mu\sigma^2_d}{P_r^{\mathrm{max}}} \leq |h_{rd}|^2 < \frac{\mu\sigma^2_d}{P_r^{\mathrm{max}}-(\mu+1)P_{\Delta}}, \\
    0,  &|h_{rd}|^2 <\frac{\mu\sigma^2_d}{P_r^{\mathrm{max}}}.
  \end{array}
\right. \notag
\end{align}
As per \eqref{PR1}, we note that R also does not transmit covert message when it cannot support the transmission from S to D (i.e., when $|h_{rd}|^2 <{\mu\sigma^2_d}/{P_r^{\mathrm{max}}}$). This is due to the fact that a transmission outage occurs when $|h_{rd}|^2 <{\mu\sigma^2_d}/{P_r^{\mathrm{max}}}$ and D will request a retransmission from S, which enables S to detect R's covert transmission with probability one if this happens. In summary, S cannot detect R's covert transmission with probability one (R could possibly transmit covert message without being detected) only when the condition $ |h_{rd}|^2 \geq {\mu\sigma^2_d}/{[P_r^{\mathrm{max}}-(\mu+1)P_{\Delta}]}$ is guaranteed. We denote this necessary condition for covert communication as $\mathbb{C}$. Considering Rayleigh fading, the cumulative distribution function (cdf) of $|h_{rd}|^2$ is given by $F_{|h_{rd}|^2}(x) = 1 - e^{-x}$ and thus the probability that $\mathbb{C}$ is guaranteed is given by
\begin{align}\label{PC}
\mathcal{P}_c = \exp\left\{-\frac{\mu\sigma^2_d}{P_r^{\mathrm{max}}-(\mu+1)P_{\Delta}}\right\}.
\end{align}
We note that $\mathcal{P}_c$ is a monotonic decreasing function of $P_{\Delta}$, which indicates that the probability that R can transmit covert message (without being detected with probability one) decreases as $P_{\Delta}$ increases. Following \eqref{Pr1_special} and noting $P_r^1 + P_{\Delta} \leq P_r^{\mathrm{max}}$, we have $P_r^{\mathrm{max}} > (\mu+1)P_{\Delta}$ and thus $0 \leq \mathcal{P}_c \leq 1$.

\section{Binary Detection at Source}

In this section, we first present the detection strategy adopted by S (i.e., Source) and then analyze the associated detection performance in terms of the false alarm and miss detection rates. Finally, the optimal detection threshold at S that minimizes the total error rate is examined.

\subsection{Binary Hypothesis Test}

In phase 2 when R transmits to D, S is to detect whether R transmits the covert message $\mathbf{x}_c$ on top of forwarding S's message $\mathbf{x}_b$ to D. In this section, we only focus on the case when $\mathbb{C}$ is guaranteed since R never transmits covert message when $\mathbb{C}$ is not met, as discussed in Section III-B. R does not transmit $\mathbf{x}_c$ in the null hypothesis $\Hnull$ while it does in the alternative hypothesis $\Halt$. Then, the received signal at S in phase 2 is given by
\begin{eqnarray}\label{yw}
\mathbf{y}_s[i]\!=\!
 \left\{ \begin{aligned}\label{ncon}
        \ &\sqrt{P_r^0}h_{rs}\mathbf{x}_r[i] + \mathbf{n}_s[i], ~~~~~~~~~~~~~~~~~~~~~~\Hnull, \\
        \ &\sqrt{P_r^1}h_{rs}\mathbf{x}_r[i] + \sqrt{P_{\Delta}}h_{rs}\mathbf{x}_c[i] \!+\! \mathbf{n}_s[i],  ~~~\Halt,
         \end{aligned} \right.
\end{eqnarray}
where $\mathbf{n}_s[i]$ is the AWGN at S with $\sigma_s^2$ as its variance.
We note that neither $P_r^0$ nor $P_r^1$ is known at S since it does not know $h_{rd}$, while the statistics of $P_r^0$ and $P_r^1$ are known since the distribution of $h_{rd}$ is publicly known. The ultimate goal of S is to detect whether $\mathbf{y}_s$ comes from $\Hnull$ or $\Halt$ in one fading block.
As proved in \cite{Shahzad2017Covert}, the optimal decision rule that minimizes the total error rate at S is given by
\begin{align}\label{decisions}
T\mathop{\gtrless}\limits_{\Honull}^{\Hoalt}\tau,
\end{align}
where $T=1/n\sum_{i=1}^{n}|\mathbf{y}_s[i]|^2$, $\tau$ is a predetermined threshold, $\Hoalt$ and $\Honull$ are the binary decisions that infer whether R transmits covert message or not, respectively. In this work, we consider infinite blocklength, i.e., $n \rightarrow \infty$. As such, we have
\begin{eqnarray}\label{Pw}
T\!=\!
 \left\{ \begin{aligned}\label{ncon}
        \ &P_r^0 |h_{rs}|^2 + \sigma_s^2, ~~~~~~~~~~~~~~~~~\Hnull, \\
        \ &P_r^1 |h_{rs}|^2  + P_{\Delta}|h_{rs}|^2 \!+\! \sigma_s^2,  ~~~~\Halt.
         \end{aligned} \right.
\end{eqnarray}

\subsection{False Alarm and Miss Detection Rates}

In this subsection, we derive S's false alarm rate, i.e., $\mathcal{P}(\Hoalt|\Hnull)$, and miss detection rate, i.e., $\mathcal{P}(\Honull|\Halt)$.

\begin{theorem}\label{theorem1}
When the condition $\mathbb{C}$ is guaranteed, the false alarm and miss detection rates at S are derived as
\begin{align}
\mathcal{P}_{FA}&=\left\{
  \begin{array}{ll}
    1,  &\tau<\sigma_s^2, \\
    1-\kappa_1, & \sigma_s^2\leq\tau\leq\rho_1,\\
    0,  &\tau>\rho_1,
  \end{array}
\right. \label{PFA}\\
\mathcal{P}_{MD}&=\left\{
  \begin{array}{ll}
    0,  &\tau<\rho_2, \\
    \kappa_2 , & \rho_2\leq\tau\leq \rho_3,\\
    1,  &\tau>\rho_3,
  \end{array}
\right. \label{PMD}
\end{align}
where
\begin{align}
&\rho_1\triangleq[{P_r^{\mathrm{max}}-(\mu+1)P_{\Delta}}]|h_{rs}|^2+\sigma_s^2, \notag \\
&\rho_2\triangleq(\mu+1)P_{\Delta}|h_{rs}|^2+\sigma_s^2, \notag \\
&\rho_3\triangleq P_r^{\mathrm{max}}|h_{rs}|^2+\sigma^2_s, \notag \\
&\kappa_1(\tau)\triangleq\mathrm{exp}\left\{\mu\sigma^2_d\left[\frac{1}{P_r^{\mathrm{max}}-(\mu+1)P_{\Delta}}-\frac{|h_{rs}|^2}{\tau-\sigma_s^2}\right]\right\}, \notag  \\
&\kappa_2(\tau)\triangleq\mathrm{exp}\left\{\mu\sigma^2_d\left[\frac{1}{P_r^{\mathrm{max}}-(\mu+1)P_{\Delta}}-\frac{|h_{rs}|^2}{\tau-\rho_2}\right]\right\}. \notag
\end{align}
\end{theorem}
\begin{IEEEproof}
Considering the maximum power constraint at R under $\Hnull$ (i.e., $P_r^0 \leq P_r^{\mathrm{max}}$) and following \eqref{PR0}, \eqref{decisions}, and \eqref{Pw}, the false alarm rate under the condition $\mathbb{C}$ is given by
\begin{align}
&\mathcal{P}_{FA}=\mathcal{P}\left[\frac{\mu\sigma^2_d}{|h_{rd}|^2}|h_{rs}|^2+\sigma_s^2\geq\tau\big{|}\mathbb{C}\right]  \\
&=\left\{
  \begin{array}{ll}
    1,  &\tau<\sigma_s^2, \\
    \frac{\mathcal{P}\left[\frac{\mu\sigma^2_d}{P_r^{\mathrm{max}}-(\mu+1)P_{\Delta}}\leq\left|h_{rd}\right|^2\leq\frac{\mu\sigma^2_d\left|h_{rs}\right|^2}
    {\tau-\sigma_s^2}\right]}{\mathcal{P}_c} , &\sigma_s^2\leq\tau\leq\rho_1,\\
    0,  &\tau>\rho_1.
  \end{array}
\right. \notag
\end{align}
Then, substituting $F_{|h_{rd}|^2}(x) = 1 - e^{-x}$ into the above equation ($h_{rs}$ is perfectly known by S and thus it is not a random variable here) we achieve the desired result in \eqref{PFA}.

We first clarify that we have $\rho_2 < \rho_3$ due to $P_r^{\mathrm{max}} > (\mu+1)P_{\Delta}$ as discussed after \eqref{PC}. Then, considering the maximum power constraint at R under $\Halt$ (i.e., $P_r^1 + P_{\Delta} \leq P_r^{\mathrm{max}}$) and following \eqref{PR1}, \eqref{decisions}, and \eqref{Pw}, the miss detection rate under the condition $\mathbb{C}$ is given by
\begin{align}
&\mathcal{P}_{MD}=\mathcal{P}\left[\left(\frac{\mu\sigma^2_d}{|h_{rd}|^2} + (1+\mu)P_{\Delta}\right)|h_{rs}|^2+\sigma_s^2<\tau\big{|}\mathbb{C}\right] \notag \\
&=\left\{
\begin{array}{ll}
    0,  &\tau<\rho_2, \\
    \frac{\mathcal{P}\left[|h_{rd}|^2\geq\frac{\mu\sigma^2_d|h_{rs}|^2}{\tau-(\mu+1)P_{\Delta}|h_{rs}|^2-\sigma_s^2}\right]}{\mathcal{P}_c}, &\rho_2\leq\tau\leq \rho_3,\\
    1,  &\tau>\rho_3.
\end{array}
\right.
\end{align}
Then, substituting $F_{|h_{rd}|^2}(x) = 1 - e^{-x}$ into the above equation we achieve the desired result in \eqref{PMD}.
\end{IEEEproof}

We note that the false alarm and miss detection rates given in Theorem~\ref{theorem1} are functions of the threshold $\tau$ and we examine how S sets the value of it in order to minimize its total error rate. Specifically, the total error rate of the detection at S is defined as
\begin{align}
\xi\triangleq\mathcal{P}_{FA} + \mathcal{P}_{MD},
\end{align}
which is used to measure the detection performance at S.

\subsection{Optimization of the Detection Threshold at Source}

In this subsection, we examine how S optimally sets the value of $\tau$ aiming to minimize $\xi$. To this end, we first determine a preliminary constraint on $P_{\Delta}$ and the bounds on the optimal $\tau$ in the following theorem.

\begin{theorem}\label{theorem2}
R's transmit power of the covert message $P_{\Delta}$ should satisfy
\begin{align}\label{P_condition}
P_{\Delta}\leq P_{\Delta}^u \triangleq P_r^{\mathrm{max}}/[2(\mu+1)]
\end{align}
in order to guarantee $\xi > 0$ and when \eqref{P_condition} is guaranteed the optimal $\tau$ ($\tau^{\ast}$) at S that minimizes $\xi$ should satisfy $\rho_2 \leq \tau^{\ast} \leq \rho_1$.
\end{theorem}
\begin{IEEEproof}
When $\rho_1< \rho_2$ that requires $P_{\Delta}> P_r^{\mathrm{max}}/[2(\mu+1)]$ as per Theorem~\ref{theorem1}, following \eqref{PFA} and \eqref{PMD}, we have
\begin{align} \label{case 2 eq}
\xi=\left\{
  \begin{array}{ll}
    1, & \tau \leq \sigma_s^2, \\
    1-\kappa_1(\tau),  & \sigma_s^2<\tau < \rho_1, \\
    0,  &\rho_1 \leq  \tau\leq\rho_2, \\
    \kappa_2(\tau),  &\rho_2<\tau<\rho_3, \\
    1,  &\tau\geq \rho_3.
  \end{array}
\right.
\end{align}
This indicates that S can simply set $\tau\in\left[\rho_1, \rho_2 \right]$ to ensure $\xi=0$ when $P_{\Delta}> P_r^{\mathrm{max}}/[2(\mu+1)]$, i.e.,
S can detect the covert transmission with probability one. As such, $P_{\Delta}$ should satisfy \eqref{P_condition} in order to guarantee $\xi >0$.

When $P_{\Delta}\leq P_r^{\mathrm{max}}/[2(\mu+1)]$, i.e., $\rho_2< \rho_1$, following \eqref{PFA} and \eqref{PMD}, we have
\begin{align} \label{case 1 eq}
\xi=\left\{
  \begin{array}{ll}
    1, & \tau \leq \sigma_s^2, \\
    1-\kappa_1(\tau),  & \sigma_s^2<\tau \leq \rho_2, \\
    1-\kappa_1(\tau)+\kappa_2(\tau),  &\rho_2<\tau<\rho_1, \\
    \kappa_2(\tau),  &\rho_1\leq\tau<\rho_3, \\
    1,  &\tau\geq\rho_3,
  \end{array}
\right.
\end{align}
due to $\rho_3 > \rho_1$. Obviously, the optimal value of $\tau$ cannot satisfy $\tau \leq \sigma_s^2$ or $\tau\geq\rho_3$.

For $\sigma_s^2<\tau \leq \rho_2$, we derive the first derivative of $\xi$ with respect to $\tau$ as
\begin{align}
\frac{\partial(\xi)}{\partial\tau}=-\frac{\mu\sigma^2_d|h_{rs}|^2}{(\tau-\sigma_s^2)^2}\kappa_1<0.
\end{align}
This demonstrates that $\xi$ is a decreasing function of $\tau$ and thus we would have $\tau^{\ast} = \rho_2$ when $\sigma_s^2<\tau \leq \rho_2$.

For $\rho_1\leq\tau<\rho_3$, we derive the first derivative of $\xi$ with respect to $\tau$ as
\begin{align}
\frac{\partial(\xi)}{\partial\tau}=\frac{\mu\sigma^2_d|h_{rs}|^2}{\left[\tau-(\mu+1)P_{\Delta}|h_{rs}|^2-\sigma_s^2\right]^2}\kappa_2>0.
\end{align}
This proves that $\xi$ is an increasing function of $\tau$ and we would have $\tau^{\ast}=\rho_1$ when $\rho_1\leq\tau<\rho_3$.

Noting that $\xi$ is a continuous function of $\tau$, we can conclude that $\tau^{\ast}$ should satisfy $\rho_2 \leq \tau^{\ast} \leq \rho_1$, no mater what is the value of $\xi$ for $\rho_2<\tau<\rho_1$.
\end{IEEEproof}

The lower and upper bounds on $\tau^{\ast}$ given in Theorem~\ref{theorem2} significantly facilitate the numerical search for $\tau^{\ast}$ at S. Then, following Theorem~\ref{theorem2} and \eqref{case 1 eq}, $\tau^{\ast}$ can be obtained through
\begin{align}\label{tau_optimal}
\tau^{\ast} = \argmin_{\rho_2 \leq \tau \leq \rho_1} [1-\kappa_1(\tau)+\kappa_2(\tau)].
\end{align}
Substituting $\tau^{\ast}$ into \eqref{case 1 eq}, we obtain the minimum value of $\xi$ as $\xi^{\ast} = 1-\kappa_1(\tau^{\ast})+\kappa_2(\tau^{\ast})$.

\section{Optimization of Effective Covert Rate}

In this section, we examine the effective covert rate achieved in the considered system subject to a covert requirement.

\subsection{Effective Covert Rate}

As discussed in Section III-B, R can only transmit the covert message without being detected by S with probability one under the condition $\mathbb{C}$. As such, a positive covert rate is only achievable under this condition. When the covert message is transmitted by R, D first decodes $\mathbf{x}_b$ and subtracts the corresponding component from its received signal $\mathbf{y}_d$ given in \eqref{YD2}. Then, the effective received signal used to decode the covert message $\mathbf{x}_c$ is given by
\begin{align}
\tilde{\mathbf{y}}_d[i]=\sqrt{P_{\Delta}}h_{rd} \mathbf{x}_c[i]+\sqrt{P_r^1}h_{rd} G \mathbf{n}_r[i] + \mathbf{n}_d[i].
\end{align}
As such, following \eqref{PR1} the SINR for $\mathbf{x}_c$ is
\begin{align}
\gamma_c = \frac{P_{\Delta}(\eta|h_{sr}|^2+1)|h_{rd}|^2}{\mu P_{\Delta}|h_{rd}|^2+(\eta|h_{sr}|^2+\mu+1)\sigma^2_d},
\end{align}
where $\eta\triangleq P_s/\sigma^2_r$. Then, the covert rate achieved by R is $R_c = \log_2(1+\gamma_c)$. We next derive the effective covert rate, i.e., averaged $R_c$ over all realizations of $|h_{rd}|^2$, in the following theorem.

\begin{theorem}\label{theorem3}
The achievable effective covert rate $\overline{R}_c$ by R is derived as a function of $P_{\Delta}$ given by
\begin{align}\label{T}
\overline{R}_c&=\frac{1}{\ln 2}\exp\left\{-\frac{\mu \sigma^2_d}{P_r^{\mathrm{max}}-(\mu+1)P_{\Delta}}\right\} \times \notag \\
&\left[\ln \left(\frac{\beta_1}{\beta_2}\right)+e^{\frac{\beta2}{\alpha2}}\mathbf{Ei}\left(-\frac{\beta_2}{\alpha_2}\right)- e^{\frac{\beta_1}{\alpha_1}}\mathbf{Ei}\left(-\frac{\beta_1}{\alpha_1}\right)\right],
\end{align}
where
\begin{align}
\beta_1&\triangleq[\eta|h_{sr}|^2+(\mu+1)](P_r^{\mathrm{max}}-P_{\Delta})\sigma^2_d, \notag \\
\beta_2&\triangleq\left\{\frac{\eta|h_{sr}|^2+(\mu+1)}{[P_r^{\mathrm{max}}-(\mu+1)P_{\Delta}]^{-1}}+\mu^2P_{\Delta}\right\}\sigma^2_d, \notag \\
\alpha_1&\triangleq P_{\Delta}[\eta|h_{sr}|^2+(\mu+1)][P_r^{\mathrm{max}}-(\mu+1)P_{\Delta}], \notag \\
\alpha_2&\triangleq\mu P_{\Delta}[P_r^{\mathrm{max}}-(\mu+1)P_{\Delta}], \notag
\end{align}
and the exponential integral function $\mathbf{Ei}(\cdot)$ is given by
\begin{align}
\mathbf{Ei}(x)=-\int_{-x}^{\infty} \frac{e^{-t}}{t} dt,~~~[x<0].
\end{align}
\end{theorem}

\begin{IEEEproof}
A positive covert rate is only achievable under the condition $\mathbb{C}$ and thus $\overline{R}_c$ is given by
\begin{align}
\overline{R}_c &=\int^{\infty}_{\frac{\mu \sigma^2_d}{P_r^{\mathrm{max}}-(\mu+1)P_{\Delta}}} R_c f(|h_{rd}|^2) d |h_{rd}|^2 \notag \\
&\overset{a}{=}\frac{1}{\ln 2}\exp\left\{-\frac{\mu \sigma^2_d}{P_r^{\mathrm{max}}-(\mu+1)P_{\Delta}}\right\} \times \notag \\
&~~~\int^{\infty}_{0}\ln\left(\frac{\beta_1+\alpha_1 x}{\beta_2+\alpha_2 x}\right)e^{-x} dx,\label{R_C_details}
\end{align}
where $\overset{a}{=}$ is achieved by exchanging variables (i.e., setting $x=|h_{rd}|^2-{\mu \sigma^2_d}/[{P_r^{\mathrm{max}}-(\mu+1)P_{\Delta}}]$).
We then solve the integral in \eqref{R_C_details} with the aid of \cite[Eq. (4.337.1)]{Gradshteyn2007table} and achieve the result given in \eqref{T}.
\end{IEEEproof}

Based on Theorem~\ref{theorem3}, we note that $\overline{R}_c$ is not an increasing function of $P_{\Delta}$, since as $P_{\Delta}$ increases $R_c$ increases but $\mathcal{P}_c$ (i.e., the probability that the condition $\mathbb{C}$ is guaranteed) decreases.

\subsection{Maximization of $\overline{R}_c$ with the Covert Constraint}

A covert transmission normally requires $\xi \geq 1 - \epsilon$, where $\epsilon \in [0,1]$ is predetermined to specify the covert constraint. In practice, it is impossible to know $\xi$ at R since the threshold $\tau$ adopted by S is unknown. In this work, we consider the worst-case scenario where $\tau$ is optimized at S to minimize $\xi$. As such, the covert constraint can be rewritten as $\xi^{\ast} \geq 1 - \epsilon$. Then, following Theorem~\ref{theorem2} the optimal value of $P_{\Delta}$ that maximizes $\overline{R}_c$ subject to this constraint can be obtained through
\begin{align} \label{optimazation_problem_1}
P_{\Delta}^{\ast} = &\argmax_{0 \leq P_{\Delta} \leq P_{\Delta}^u} \overline{R}_c \\ \nonumber
&\text {s.t.} ~~~~\xi^{\ast} \geq 1-\epsilon.
\end{align}

We note that this is a two-dimensional optimization problem that can be solved by efficient numerical searches. Specifically, for each given $P_{\Delta}$, $\xi^{\ast}$ should be obtained based on \eqref{tau_optimal} where $\tau^{\ast}$ is also numerically searched. We note that the numerical search of $P_{\Delta}^{\ast}$ and $\tau^{\ast}$ is efficient since their lower and upper bounds are explicitly given.

\begin{figure}
\centering
\includegraphics[width=2.7in, height=2.3in]{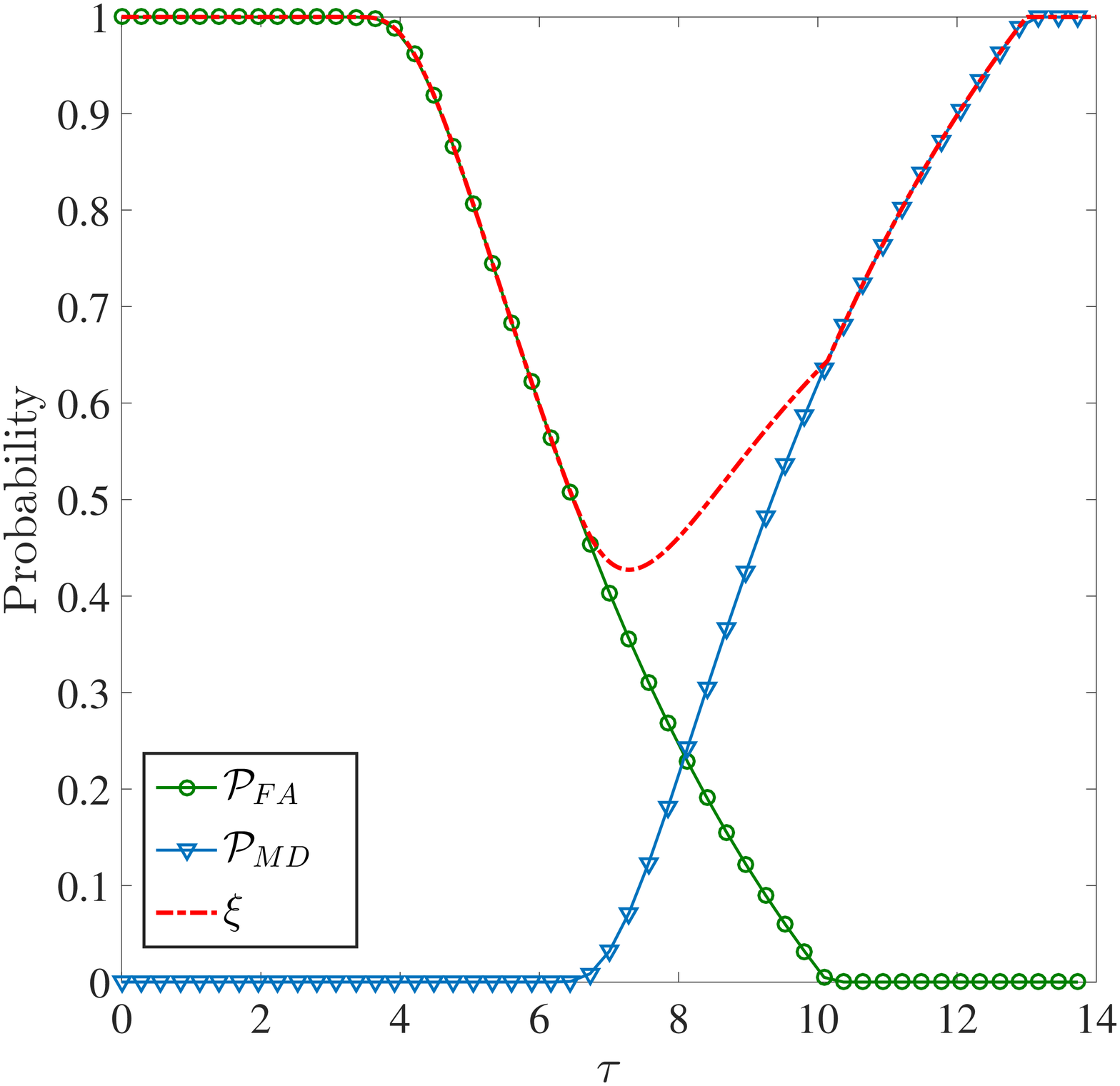}\\
\caption{$\mathcal{P}_{FA}$, $\mathcal{P}_{MD}$, and $\xi$ versus different values of the threshold $\tau$, where $P_s=P_r^{\mathrm{max}}=10$~dB, $\sigma_s^2=\sigma_r^2=\sigma^2_d=0$~dB, $P_{\Delta}=0.5$, $R_{sd}=1$, and $|h_{sr}|^2=|h_{rs}|^2=1$.}
\label{fig2}
\end{figure}

\section{Numerical Results}
In this section, we first examine the detection performance at S (i.e., Source) under the condition $\mathbb{C}$. Then, the impact of some system parameters on the
achievable effective covert rate subject to a specific covert constraint is investigated.

In Fig.~\ref{fig2}, we plot the false alarm rate $\mathcal{P}_{FA}$, miss detection rate $\mathcal{P}_{MD}$, and total error rate $\xi$ versus the threshold $\tau$, in which the adopted system parameters guarantee condition $\mathbb{C}$ and $P_{\Delta}\leq P_{\Delta}^u$. As expected, we observe that $\xi > 0$ due to the guaranteed condition $\mathbb{C}$ and $P_{\Delta}\leq P_{\Delta}^u$, which means that covert transmission is possible (not being detected with probability one) under this condition. We observe that the minimum value of $\xi$ is achieved when $\rho_2 \leq \tau \leq \rho_1$, which verifies the correctness of our Theorem~\ref{theorem2}.

\begin{figure}
\centering
\includegraphics[width=2.6in, height=2.2in]{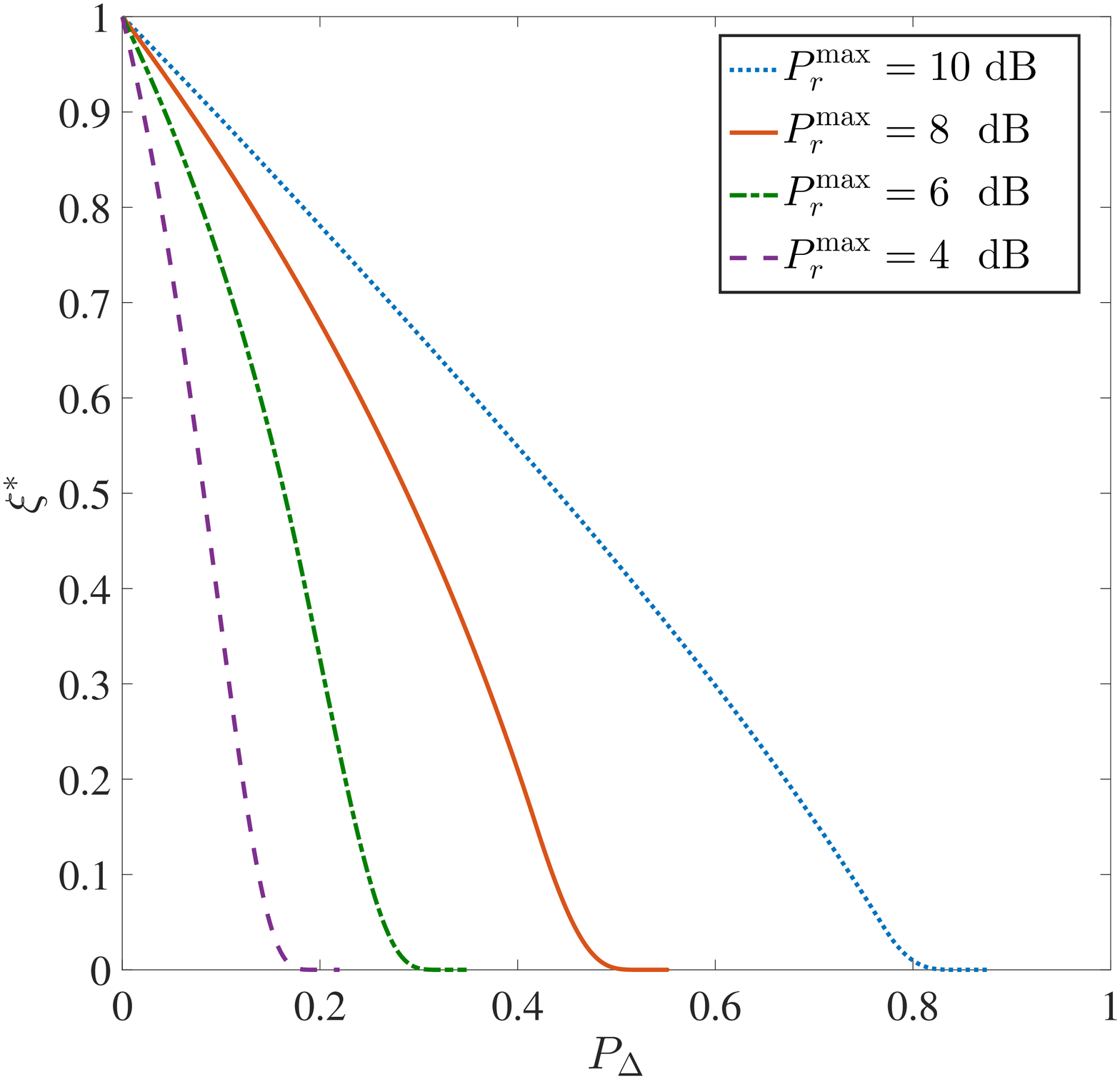}\\
\caption{$\xi^{\ast}$ versus $P_{\Delta}$ with different value of $R_{sd}$, where $P_{s}=10$~dB, $\sigma^2_r=\sigma^2_d=0$~dB, $R_{sd}=1$, and $|h_{sr}|^2=|h_{rs}|^2=1$.}
\label{fig3}
\end{figure}

In Fig.~\ref{fig3}, we plot the minimum total error rate $\xi^{\ast}$ versus the covert transmit power $P_{\Delta}$, which is achieved through searching the optimal threshold $\tau^{\ast}$ as per \eqref{tau_optimal}. In this figure, we first observe that $\xi^{\ast}$ monotonically decreases as $P_{\Delta}$ increases, which demonstrates that the covert transmission becomes easier to be detected when more power is used. As such, the covert constraint $\xi^{\ast} \geq 1-\epsilon$ determines a maximum possible  value of $P_{\Delta}$, which is significantly less than $P_{\Delta}^u$ since we have $\xi = 0$ when $P_{\Delta} = P_{\Delta}^u$ but we normally require $\xi > 0.5$ in practice \cite{ShihaoYan2017Covert}. This can facilitate the search of the optimal value of $P_{\Delta}$ as per \eqref{optimazation_problem_1} by significantly reducing the feasible region of $P_{\Delta}$. We also observe that $\xi^{\ast}$ increases as $P_r^{\mathrm{max}}$ increases. This shows that covert transmission becomes easier (i.e., the detection probability of covert transmission at S $1-\xi^{\ast}$ decreases) as the desired performance of the normal transmission increases (i.e., the transmission outage probability decreases as $P_r^{\mathrm{max}}$ increases for a fixed $R_{sd}$).

\begin{figure}
\centering
\includegraphics[width=2.6in, height=2.2in]{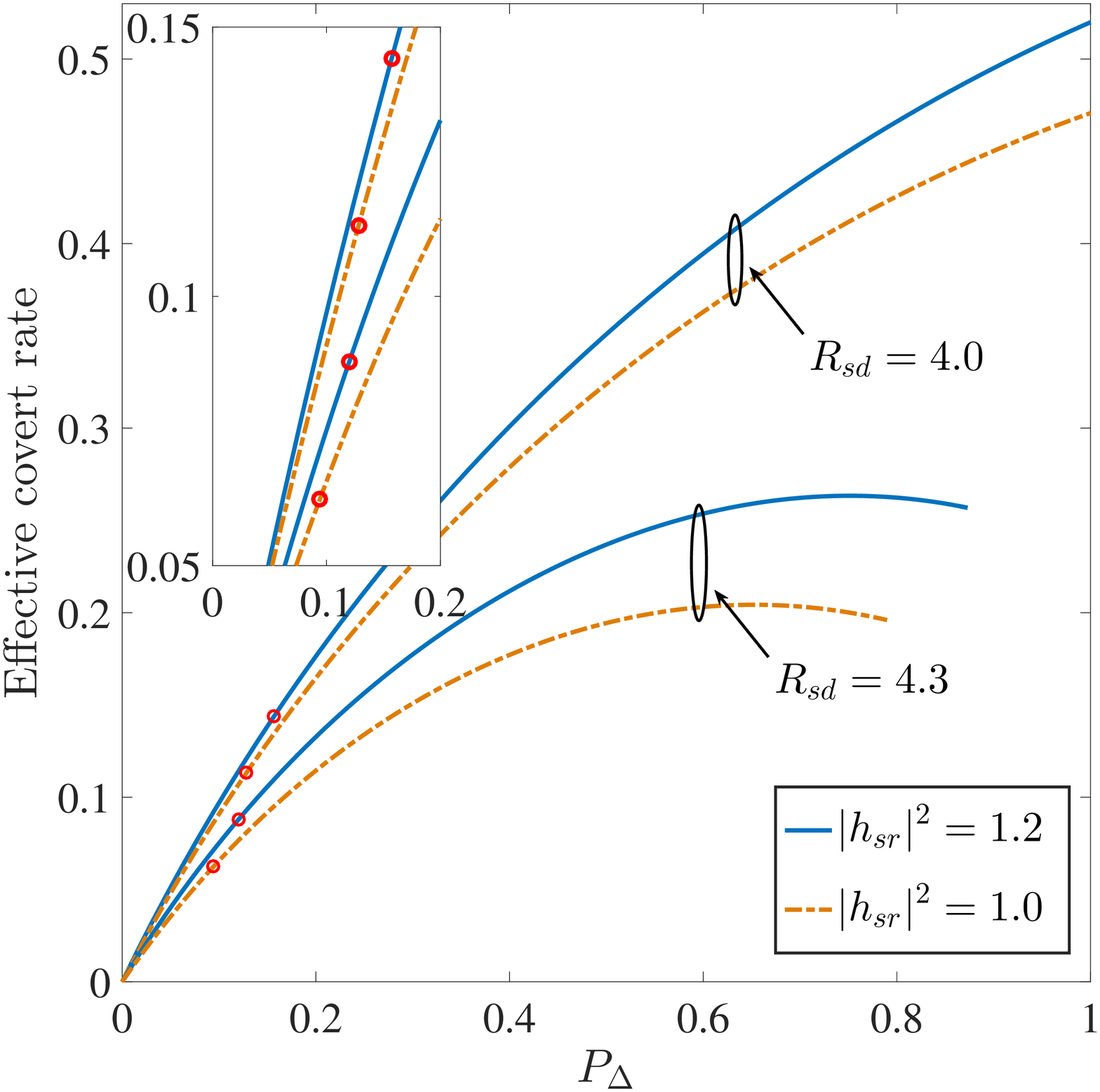}\\
\caption{$\overline{R}_c$ versus $P_{\Delta}$ with different value of $|h_{sr}|^2$, where $P_{s}=P_r^{\mathrm{max}}=30$ dB, $\sigma^2_r=\sigma^2_d=0$ dB, and $\epsilon = 0.1$ ($\epsilon $ is only for the red circles).}
\label{fig4}
\end{figure}

In Fig.~\ref{fig4}, we plot the effective covert rate $\overline{R}_c$ versus $P_{\Delta}$, in which we also show the maximum possible value of $P_{\Delta}$ determined by the covert constraint $\xi^{\ast} \geq 1-\epsilon$ (denoted by $P_{\Delta}^{\epsilon}$ and marked by red circle in this figure). We first observe that $\overline{R}_c$ may not be a monotonically increasing function of $P_{\Delta}$ without the constraint $\xi^{\ast} \geq 1- \epsilon$. This is due to the fact that as $P_{\Delta}$ increases the probability to guarantee the condition $\mathbb{C}$ (i.e., $\mathcal{P}_c$) decreases while the covert rate $R_c$ increases. In addition, we observe that $\overline{R}_c$ without $\xi^{\ast} \geq 1- \epsilon$ increases as $|h_{sr}|^2$ increases. This is due to the fact that as $|h_{sr}|^2$ increases $\mu$ as given in \eqref{mu} decreases, which leads to that $\mathcal{P}_c$ increases, i.e., the probability that R can conduct covert transmission increases (although the covert rate $R_c$ does not change). Finally, we observe that  $P_{\Delta}^{\epsilon}$ increases as well when $|h_{sr}|^2$ increases.  As such, following the last two observations we can conclude that the achievable effective covert rate with the constraint $\xi^{\ast} \geq 1- \epsilon$ increases as  $|h_{sr}|^2$ increases. Intuitively, this is due to that
as $|h_{sr}|^2$ increases R has a higher chance to support the transmission of $\mathbf{x}_b$ and perform covert transmission, resulting in that from S's point of view the possible transmit power range of R used to transmit $\mathbf{x}_b$ increases (i.e., transmit power uncertainty increases).

\section{Conclusion}
This work examined covert communication in one-way relay networks over Rayleigh fading channels. Specifically, we analyzed S's detection limit of the covert transmission from R to D in terms of the total error rate. We also determined the maximum achievable effective covert rate subject to $\xi^{\ast} \geq 1-\epsilon$. Our examination shows that covert communication in the considered relay networks is feasible and a tradeoff between the achievable effective covert rate and R's performance of aiding the transmission from S to D exists.

\section*{Acknowledgments}
This work was supported by the Australian Research Council's Discovery Projects (DP150103905), the National Natural Science Foundation of China (No. 61472190), the Open Research Fund of National Key Laboratory of Electromagnetic Environment, China Research Institute of Radiowave Propagation (No. 201500013), and the open research fund of National Mobile Communications Research Laboratory, Southeast University, China (No. 2013D02).

\end{document}